\documentclass{iau_FM}
\usepackage{graphicx}
\usepackage{booktabs} % for much better looking tables
\usepackage{amsmath}

\usepackage{natbib,twoopt}
\usepackage{color}
\usepackage[breaklinks=true,draft]{hyperref} %% to avoid \citeads line fills
\title[FM 8.~~Disk-Halo Interaction of M51] %% give here short title %%
{The Magnetized Disk-Halo Transition Region of M51
}

\author[Kierdorf et al. ]   %% give here short author list %%
{M. Kierdorf$^1$,
  S. A. Mao$^1$, A. Fletcher$^2$, R. Beck$^1$, M. Haverkorn$^3$, A. Basu$^4$, F. Tabatabaei$^5$, \and J. Ott$^6$
 }

\affiliation{$^1$Max-Planck-Institut f\"ur Radioastronomie, Auf dem H\"ugel 69, 53121, Bonn, Germany \\ 
email: {\tt kierdorf@mpifr-bonn.mpg.de}\\
$^2$School of Mathematics and Statistics, Herschel Building, Newcastle University, NE1 7RU U.K.\\
$^3$Department of Astrophysics/IMAPP, Radboud University Nijmegen; P.O. Box 9010, 6500 GL Nijmegen, Netherlands\\
$^4$Fakult\"at f\"ur Physik, Universit\"at Bielefeld, Universit\"atsstr. 25, 33615 Bielefeld\\
$^5$Instituto de Astrof\'isica de Canarias, San Crist\'obal de La Laguna Santa Cruz de Tenerife, Spain\\
$^6$National Radio Astronomy Observatory, 1003 Lopezville Road, Socorro, NM 87801, USA\\
}

\pubyear{2018}
\jname{Astronomy in Focus, Volume 1} 
\editors{Piero Benvenuti, ed.}
\begin{document}

\maketitle

\begin{abstract}
An excellent laboratory for studying large scale magnetic fields is the grand design face-on spiral galaxy M51. Due to wavelength-dependent Faraday depolarization, linearly polarized synchrotron emission at different radio frequencies gives a picture of the galaxy at different depths: Observations at L-band (1\,--\,2\,GHz) probe the halo region while at C- and X-band (4\,--\,8\,GHz) the linearly polarized emission probe the disk region of M51. We present new observations of M51 using the Karl G. Jansky Very Large Array (VLA) at S-band (2\,--\,4\,GHz), where previously no polarization observations existed, to shed new light on the transition region between the disk and the halo. 
We discuss a model of the depolarization of synchrotron radiation in a multilayer magneto-ionic medium and compare the model predictions to the multi-frequency polarization data of M51 between 1\,--\,8\,GHz. The new S-band data are essential to distinguish between different models. Our study shows that the initial model parameters, i.e. the total regular and turbulent magnetic field strengths in the disk and halo of M51, need to be adjusted to successfully fit the models to the data.

\end{abstract}

\firstsection % if your document starts with a section,
\section{Introduction}
\begin{figure}[t]
 \vspace*{-0.25 cm}
\centering
\includegraphics[width=4in]{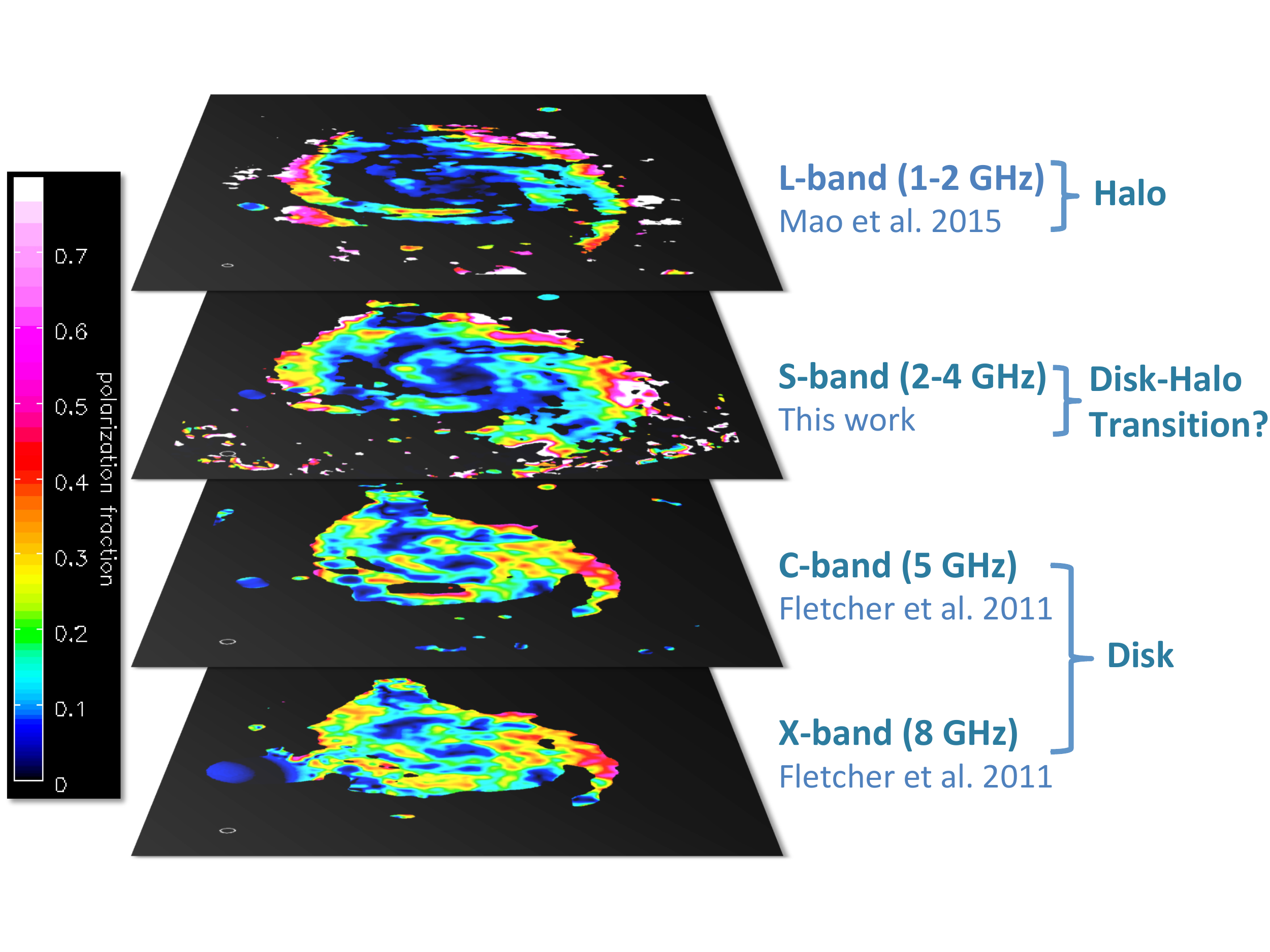} 
\vspace*{-0.8 cm}
 \caption{Observed degree of polarization of M51 at different frequencies. All images have the same color scale and are smoothed to the same resolution of 15 arcsec (which corresponds to about 550\,pc at the distance of M51). Note that the total intensity images used to calculate the degree of polarization were not corrected for thermal emission.}
 \label{fig:PD_3D}
\end{figure}

M51 is a nearby face-on grand design spiral galaxy. 
Modern radio interferometers with high spatial resolution allow us to probe detailed structures of the galaxy in total intensity and linear polarization. 
Investigating depolarization effects of linearly polarized synchrotron emission at different wavelengths is a powerful tool to put constraints on the magneto-ionic properties of the interstellar medium (ISM) in galaxies. 
One effect which can cause depolarization 
is Faraday rotation in magnetized thermal plasma which rotates the plane of linear polarization of an electro-magnetic wave by an angle proportional to $\lambda^2$. The proportionality constant is called the rotation measure (RM) and is measured in the unit of rad\,m$^{-2}$. 
By comparing the observed degree of polarization as a function of wavelength with models of depolarization, one can investigate the underlying magnetic field properties (e.g. the regular and turbulent magnetic field strengths in the ISM). 
In the ISM of spiral galaxies, cosmic rays (CRs) as well as thermal electrons are mixed with magnetic fields in the same spatial volume. This causes emission of synchrotron radiation and Faraday rotation at the same locations. In such a case the polarization plane experiences \textit{differential Faraday rotation} which results in a sinc-function variation of the fractional polarization with $\lambda^2$ \citep{1966MNRAS.133...67B}. Therefore, at different frequencies one can probe the linearly polarized emission of a face-on galaxy at different physical depths: at high radio frequencies the polarized signal from the disk of the galaxy experiences low Faraday depolarization whereas at low radio frequencies, the polarized signal from the disk is almost completely depolarized.
Figure \ref{fig:PD_3D} shows the observed degree of polarization of M51 at frequencies between 1\,--\,8\,GHz at the same angular resolution and the same color scale. 
One can see that the degree of polarization decreases with increasing wavelength (from bottom to top). Especially at L-band (top panel of Figure \ref{fig:PD_3D}) the central region of M51 is strongly depolarized. A detailed description of the data reduction and analysis and the full result of this work will be included in a forthcoming paper (Kierdorf et al., in prep.).

\vspace*{-0.5cm}
\section{M51's ``unknown'' polarization layer}
Polarization studies of M51 shows that different configurations of the regular magnetic field exists in the disk and in the halo (e.g. \citet{Fletcher11}). 
According to \citet{Fletcher11}, the regular field in the disk is best described by a superposition of two azimuthal modes (axisymmetric plus quadrisymmetric), %($m\,=\,0$ and 2)
 whereas the halo field has a strong bisymmetric azimuthal mode. The clear difference in the magnetic field configuration between the disk and the halo of M51 is still poorly understood. 
A better understanding will come from observations of the transition region between the disk and the halo. To investigate the ``unknown'' polarized layer between the disk and the halo we observed M51 in S-band (2\,--\,4\,GHz, 7.5 - 15\,cm) 
where no polarization data existed previously. 
We used the Karl G. Jansky Very large Array (VLA) in Soccoro, New Mexico operated by the National Radio Astronomy Observatory (NRAO) which provides large antenna separation and wideband receivers resulting in high spatial resolution and wide frequency coverage with high sensitivity. 
Our new broadband S-band polarization data fills the gap between data observed with the VLA at L-band (1-2\,GHz) by \citet{2015ApJ...800...92M}, and C-band (4.85\,GHz) and X-band (8.35\,GHz) by \citet{Fletcher11}. With this combined high quality and broad frequency coverage data set we are able to investigate the magneto-ionic properties in different layers of M51.  

\vspace*{-0.5cm}
\section{Faraday depolarization in a multi-layer magneto-ionic medium}
\begin{figure}[htbp]
\centering
\vspace*{-0.25cm}
\hspace*{-0.2 cm} \includegraphics[width=2.6in]{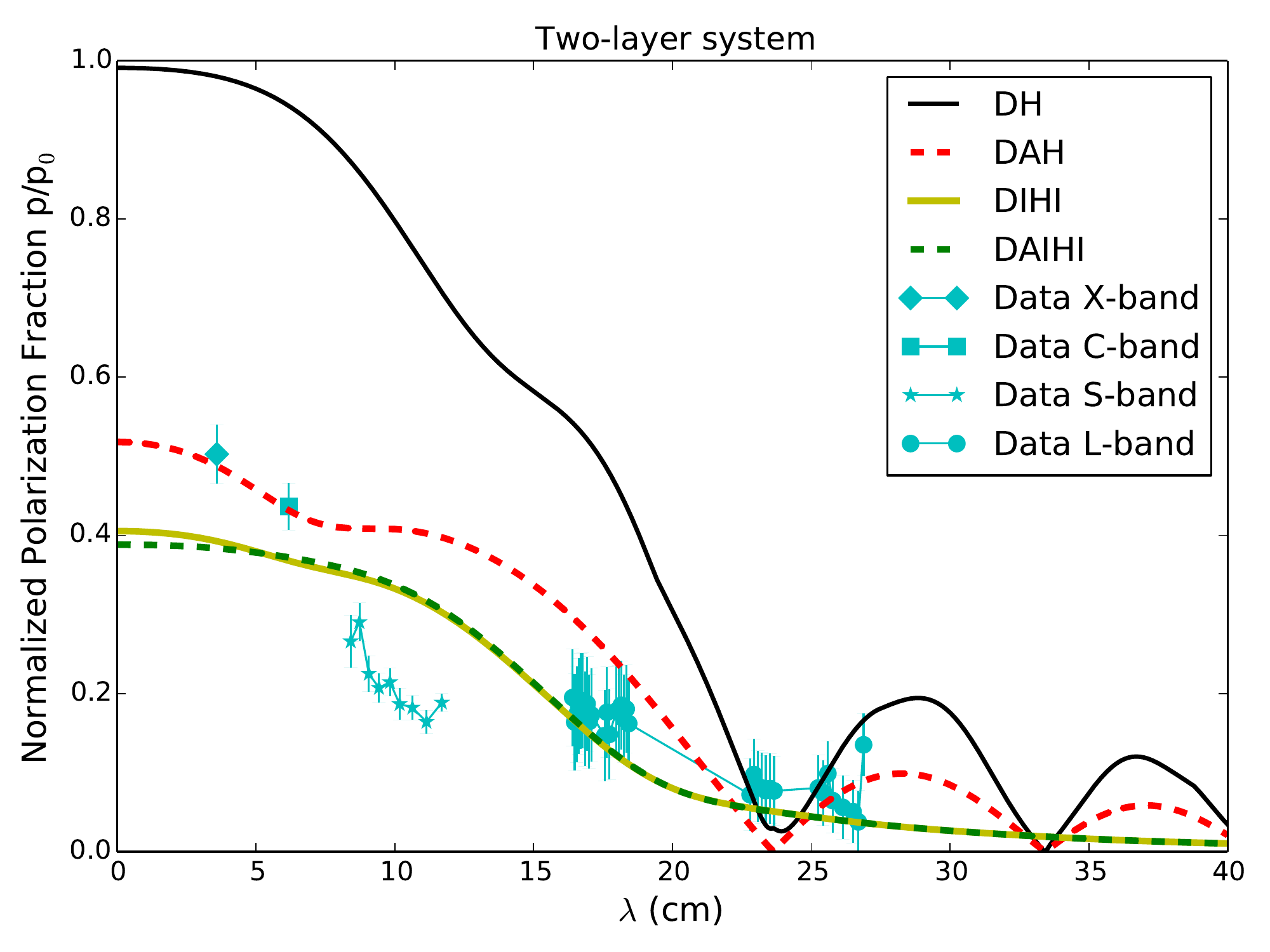}\includegraphics[width=2.6in]{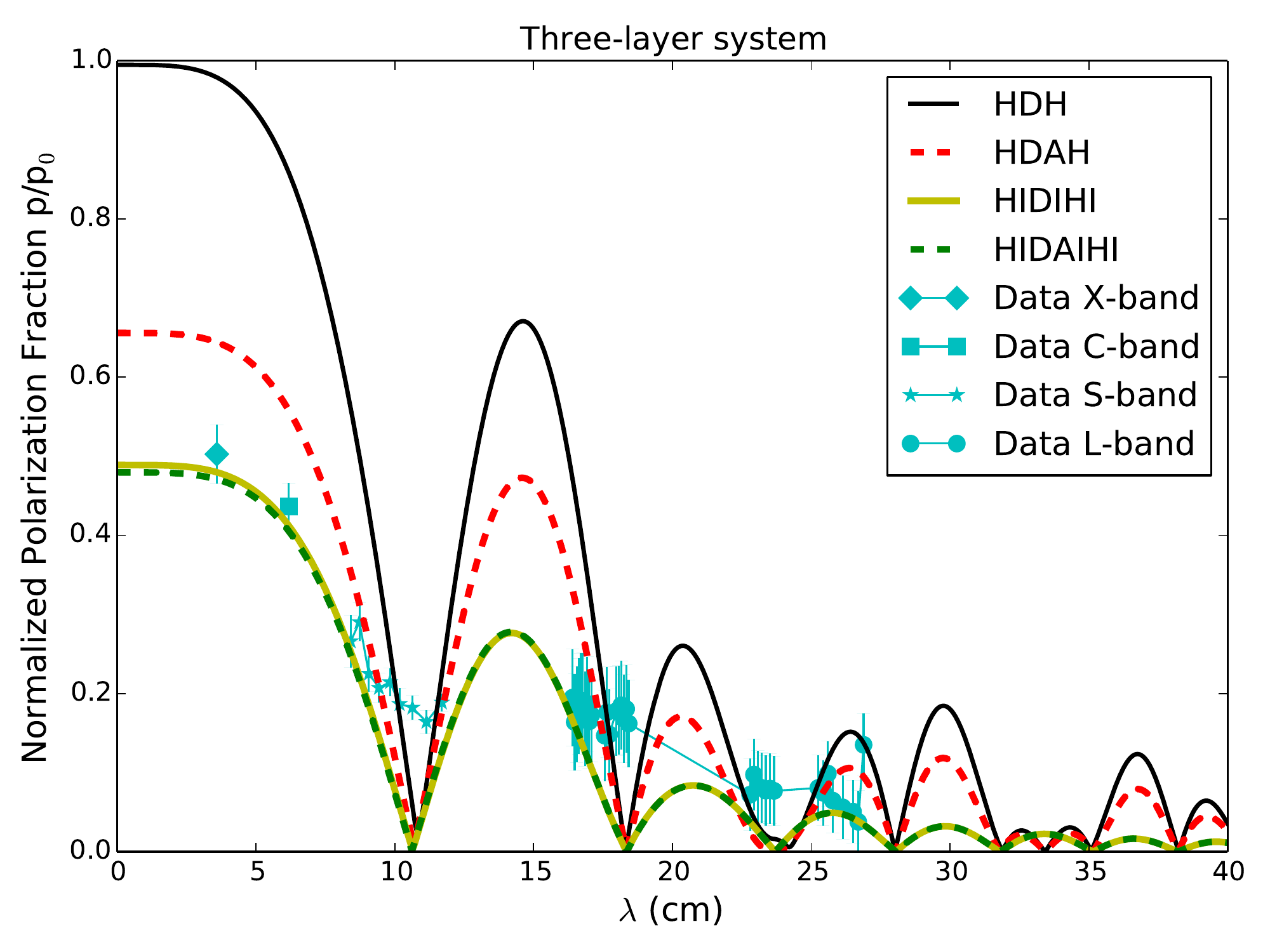}
\vspace*{-0.2cm}
\caption{Depolarization models from \citet{Shneider14} for a two-layer system (left panel) 
and a three-layer system (right panel) 
in M51 plotted together with the observed degree of polarization at multiple wavelengths. 
All model profiles featured have been constructed from a set of the following parameters: a total regular magnetic field strength of 5\,$\mu$G in the disk and halo, a disk turbulent random field of 14\,$\mu$G, and a halo turbulent random field of 4\,$\mu$G and a thermal electron density of 0.11\,cm$^{-3}$ and 0.01\,cm$^{-3}$ in the disk and halo, respectively.
For nomenclature and description of the model types appearing in the legend we refer to the text. 
} 
\label{fig:shneider_data}
  \end{figure}

\citet{Shneider14} developed a model of the depolarization of synchrotron radiation in a multilayer magneto-ionic medium. They developed model predictions for the degree of polarization as a function of wavelength for a two-layer system with a disk and a halo and a three-layer system with a far-side halo, a disk and a near-side halo.  
The model includes differential Faraday rotation caused by regular magnetic fields and internal Faraday dispersion due to random magnetic fields. In the case of a three-layer system, the near and far-side halo have identical properties. Figure \ref{fig:shneider_data} shows the model predictions of the normalized degree of polarization ($p/p_0$) for a two-layer (left panel) and three-layer (right panel) system, respectively. $p_0$ is the intrinsic degree of polarization which is assumed to be 70\,\% corresponding to the theoretical injection spectrum for electrons accelerated in supernova-remnants with a synchrotron spectral index of $\alpha_{\textrm{syn}}\,=\,-0.5$ \citep{Shneider14}). The nomenclature of the different models is as follows: `D' and `H' stands for regular fields in the disk and halo, respectively. `I' and `A' denotes isotropic and anisotropic turbulent fields where the first one is for the disk and the second for the halo. 
The observed degrees of polarization at X-band, C-band, S-band and L-band are also shown. The total and polarized intensity was integrated in a sector with an azimuthal angle centered at 100$^\circ$ and an opening angle of 20$^\circ$ and radial boundaries 2.4\,--\,3.6\,kpc\footnote{To obtain the non-thermal total flux density at this location, we assumed a thermal fraction $f^{\text{th}}_{\nu_0}$ of 9\,\% at $\nu_0=\,3$\,GHz \citep{2017ApJ...836..185T} and extrapolated the non-thermal flux densities at frequency $\nu$ via $S_{\nu}=f^{\text{th}}_{\nu_0}\left(\frac{\nu}{\nu_0}\right)^{-0.1}S_{\nu_0}$.}. 
With only the data points at 3.6\,cm, 6.2\,cm, and L-band (15\,--\,30\,cm) it is not possible to assess if a two-layer or three-layer system is more likely for M51. 
Since the model predictions strongly differ within the wavelength range of S-band (7.5\,-15\,cm), our new S-band data are essential to distinguish between the different systems. 

By comparing the observed degree of polarization to the models, one can directly rule out models with only regular magnetic fields in the disk and halo (DH) since the observed data deviates most from those model predictions. 
However, it appears that none of the model predictions with the parameters given in \citet{Shneider14} are in agreement with the observed data at S-band. For the two-layer system, the data points deviates from the model whereas for the three-layer case, some data points are well reproduced by the model predictions but the model drops to zero at $\lambda\approx11$\,cm which is clearly ruled out by the observed data.

\begin{figure}[htbp]
\centering
\vspace*{-0.25cm}
\hspace*{-0.2 cm} 
\includegraphics[width=2.6in]{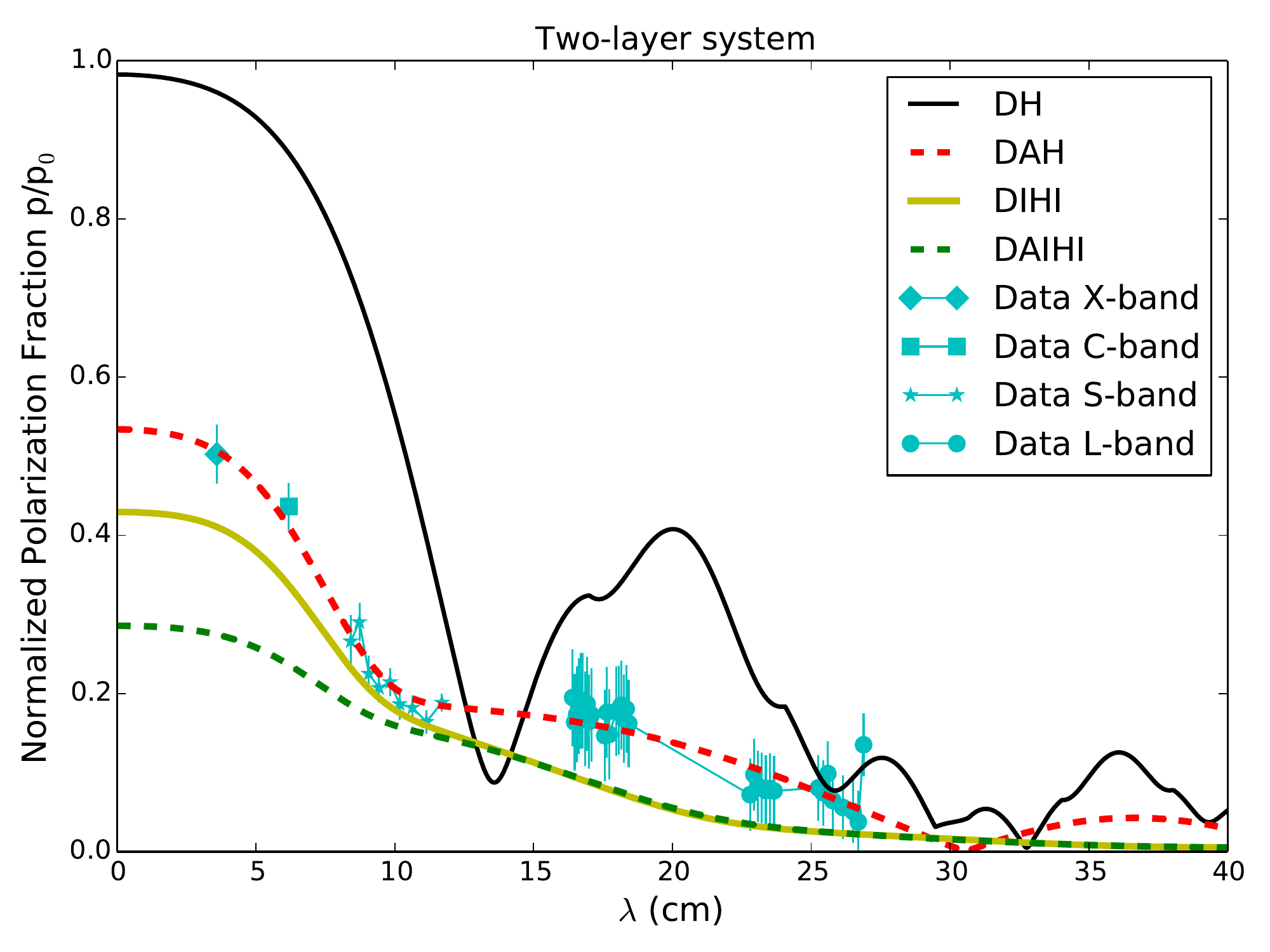}\includegraphics[width=2.6in]{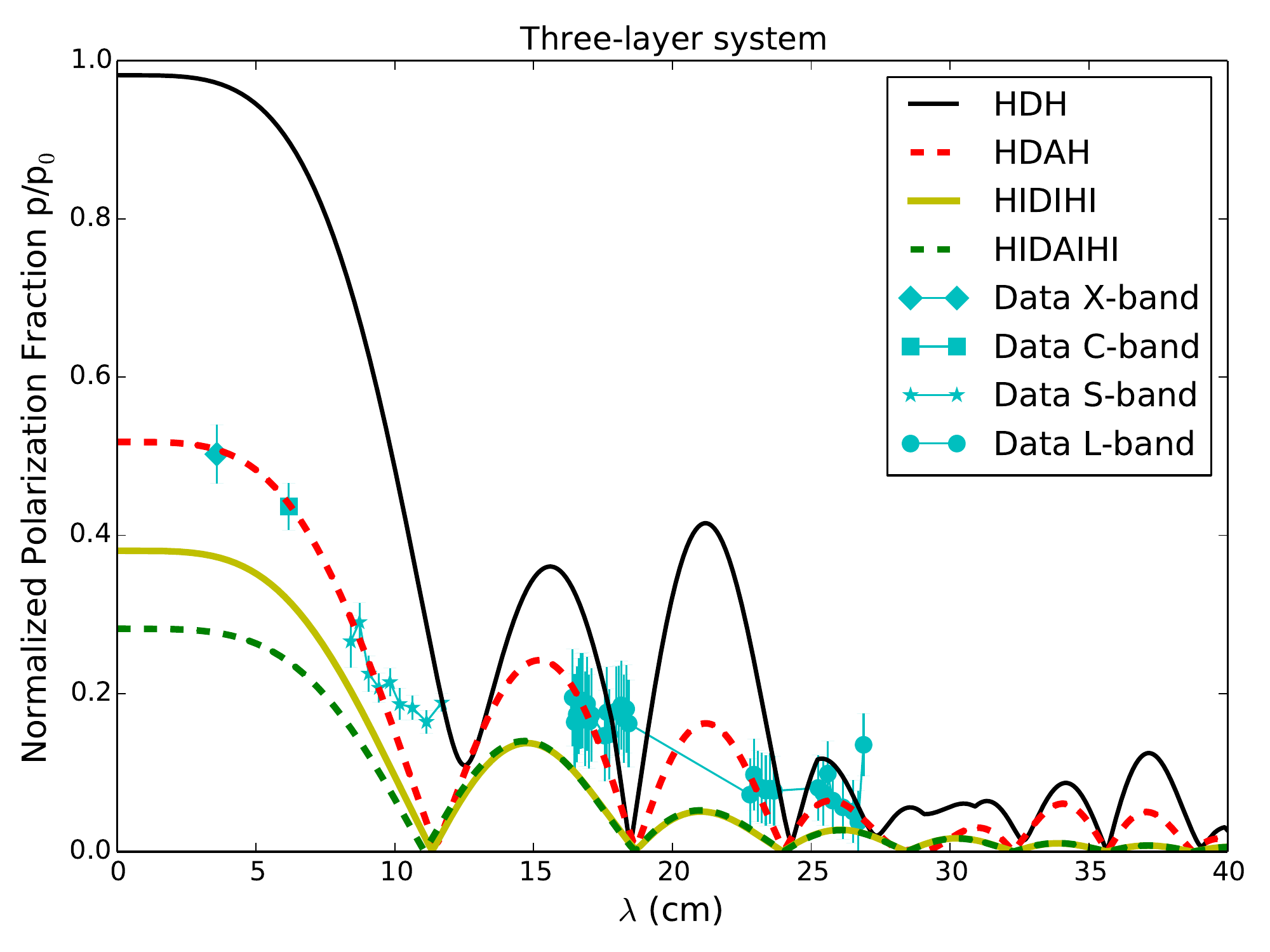}
\vspace*{-0.2cm}
\caption{``Best-fit'' of the model DAH for a two-layer system (left panel) and a three-layer system (right panel) in M51 to the observed degree of polarization at multiple wavelengths. The ``best fit'' magnetic field strengths are 10\,$\mu$G and 3\,$\mu$G for the total regular field in the disk and in the halo for the two-layer and three-layer system, respectively. The total random magnetic field strengths in the disk amounts to 14\,$\mu$G and 16\,$\mu$G for the two - and three-layer system, respectively. For the electron density a value of 0.07\,cm$^{-3}$ and 0.01\,cm$^{-3}$ in the disk and halo fits the data best in both layer systems.
} 
\label{fig:shneider_fit}
  \end{figure}

To investigate the influence of different total magnetic field strengths on the degree of polarization, we developed an interactive tool which allows one to produce model predictions for a range of total regular and turbulent magnetic field strengths in the disk and halo simultaneously. 
Figure \ref{fig:shneider_fit} shows the ``best fit'' of the model DAH (red dashed line) with regular magnetic fields in disk and halo and anisotropic turbulent magnetic fields in the disk. We explored visually whether any reasonable combination of the free parameters can reproduce the observed degree of polarization. These ``best fit'' magnetic field strengths and electron densities are listed in the caption of Figure \ref{fig:shneider_fit} and are all physically plausible values.
For the three-layer system it is not possible to lift up the zero points in the model by changing any of the free parameters. Therefore, this three-layer model 
can be ruled out. In other words, we do not detect any polarized emission from the far side halo.

\vspace*{-1cm}
\section{Future Work}
We show that the comparison of the observed degree of polarization to the wavelength-dependent depolarization models is a powerful tool to put constraints on the magnetic field strengths and thermal electron density in different regions of the galaxy. 
As a next step, we will apply the same method to other sectors with different azimuthal angles and radiuses in M51. 
Consistency between fits to different sectors is a strong indication that the model is physically meaningful. Further, since the different models distinguish between isotropic and anisotropic turbulent magnetic fields, by comparing the observed degree of polarization in different regions of the galaxy with the model predictions, we can investigate turbulent magnetic field configurations in different locations in M51. 

\vspace*{-0.5cm}
\bibliography{references}{}

\begin{thebibliography}{}
\makeatletter
\relax
\def\mn@urlcharsother{\let\do\@makeother \do\$\do\&\do\#\do\^\do\_\do\%\do\~}
\def\mn@doi{\begingroup\mn@urlcharsother \@ifnextchar [ {\mn@doi@}
  {\mn@doi@[]}}
\def\mn@doi@[#1]#2{\def\@tempa{#1}\ifx\@tempa\@empty \href
  {http://dx.doi.org/#2} {doi:#2}\else \href {http://dx.doi.org/#2} {#1}\fi
  \endgroup}
\def\mn@eprint#1#2{\mn@eprint@#1:#2::\@nil}
\def\mn@eprint@arXiv#1{\href {http://arxiv.org/abs/#1} {{\tt arXiv:#1}}}
\def\mn@eprint@dblp#1{\href {http://dblp.uni-trier.de/rec/bibtex/#1.xml}
  {dblp:#1}}
\def\mn@eprint@#1:#2:#3:#4\@nil{\def\@tempa {#1}\def\@tempb {#2}\def\@tempc
  {#3}\ifx \@tempc \@empty \let \@tempc \@tempb \let \@tempb \@tempa \fi \ifx
  \@tempb \@empty \def\@tempb {arXiv}\fi \@ifundefined
  {mn@eprint@\@tempb}{\@tempb:\@tempc}{\expandafter \expandafter \csname
  mn@eprint@\@tempb\endcsname \expandafter{\@tempc}}}

\bibitem[\protect\citeauthoryear{{Burn}}{{Burn}}{1966}]{1966MNRAS.133...67B}
{Burn} B.~J.,  1966, \mn@doi [MNRAS] {10.1093/mnras/133.1.67}, \href
  {http://adsabs.harvard.edu/abs/1966MNRAS.133...67B} {133, 67}

\bibitem[\protect\citeauthoryear{{Fletcher}, {Beck}, {Shukurov}, {Berkhuijsen}
  \& {Horellou}}{{Fletcher} et~al.}{2011}]{Fletcher11}
{Fletcher} A.,  {Beck} R.,  {Shukurov} A.,  {Berkhuijsen} E.~M.,   {Horellou}
  C.,  2011, \mn@doi [MNRAS] {10.1111/j.1365-2966.2010.18065.x}, \href
  {http://adsabs.harvard.edu/abs/2011MNRAS.412.2396F} {412, 2396}

\bibitem[\protect\citeauthoryear{{Mao}, {Zweibel}, {Fletcher}, {Ott}  \&
  {Tabatabaei}}{{Mao} et~al.}{2015}]{2015ApJ...800...92M}
{Mao} S.~A.,  {Zweibel} E.,  {Fletcher} A.,  {Ott} J.,   {Tabatabaei} F.,
  2015, \mn@doi [Astrophysical Journal] {10.1088/0004-637X/800/2/92}, \href
  {http://adsabs.harvard.edu/abs/2015ApJ...800...92M} {800, 92}

\bibitem[\protect\citeauthoryear{{Shneider}, {Haverkorn}, {Fletcher}  \&
  {Shukurov}}{{Shneider} et~al.}{2014}]{Shneider14}
{Shneider} C.,  {Haverkorn} M.,  {Fletcher} A.,   {Shukurov} A.,  2014, \mn@doi
  [Astronomy \& Astropysics] {10.1051/0004-6361/201423470}, \href
  {http://adsabs.harvard.edu/abs/2014A%26A...567A..82S} {567, A82}

\bibitem[\protect\citeauthoryear{{Tabatabaei} et~al.,}{{Tabatabaei}
  et~al.}{2017}]{2017ApJ...836..185T}
{Tabatabaei} F.~S.,  et~al., 2017, \mn@doi [Astrophysical Journal]
  {10.3847/1538-4357/836/2/185}, \href
  {http://adsabs.harvard.edu/abs/2017ApJ...836..185T} {836, 185}

\makeatother
\end{thebibliography}
\bibliographystyle{mnras}

\end{document}